# Understanding of droplet dynamics and deposition area in electrospraying process: Modeling and experimental Approaches

Zhuoying Jiang, Xiong (Bill) Yu[*]

Department of Civil Engineering, Case Western Reserve University, Cleveland, Ohio 44106, USA

[*]Correspondence: xxy21@case.edu

**Abstract**

Electrospraying is a widely-used technique for generating microspherical droplets in biomedical and chemical applications and considered as an effective approach for the deposition on substrate. However, studies on effects of controllable parameters on deposition area of electrosprayed droplets have not been reported yet. In this study, a simplified two-dimensional model is developed to study the dynamic process of droplets and electrosprayed area. The effects of distance between the needle tip and collector, and syringe feed rate on the size of sprayed area are quantified. Experiments have been conducted to validate the simulation results. Both modeling and experimental data demonstrate that the diameter of sprayed area increases with increased distance between the tip and collector as well as feed rate. This fundamental understanding should contribute to a wise choice of controllable parameters to achieve an efficient utilization of electrosparyed droplets or substrate for real applications.

*Nomenclature*

| | | |
|---|---|---|
| $E$ | electric field | V m$^{-1}$ |
| $Q$ | syringe feed rate | mL h$^{-1}$ |
| $K$ | electrical conductivity of liquid | µS cm$^{-1}$ |
| $\mu_{liquid}$ | viscosity of liquid | Pa·s |
| $\mu_{air}$ | viscosity of air | Pa·s |
| $\gamma$ | surface tension of liquid and air | mN m$^{-1}$ |
| $\beta$ | relative permittivity of liquid | |
| $\varepsilon_0$ | vacuum permittivity | F m$^{-1}$ |
| $\rho_{air}$ | density of air | m$^3$ kg$^{-1}$ |
| $\rho$ | density of a droplet | m$^3$ kg$^{-1}$ |
| $d$ | diameter of a droplet | m |
| $q_{max}$ | maximum charge of a droplet | C |
| $v$ | velocity of a droplet | m s$^{-1}$ |
| $m$ | mass of a droplet | kg |
| $V$ | volume of a droplet | m$^3$ |
| $\boldsymbol{r}$ | position of the droplet | m |

## 1. Introduction

Electrospraying approach has been widely used to deposit droplets in biomedical and chemical applications [1]. In the electrospraying process, the liquid droplets include metal ion solution [2-4], polymer solution [5-7] and ceramic suspension [8-13]. The electrosprayed droplets bring in various applications: drug delivery [5,6], ceramic films [11,12], lithium batteries [2-4], fuel cells [14,15] and gas sensors [16-19]. Various merits of electrospraying have been demonstrated: (1) simple device setup, (2) low cost and (3) capable of producing small droplets [1]. The size of the droplets is usually in a micro-scale, and can be even smaller than 100 nanometers [6,20]. Homogeneously distributed droplets are feasible to be fabricated, and this is because electrostatic repulsive forces induced among the charged droplets, thus agglomerations of droplets are avoided.

Up to date, efforts have been contributed to the application of electrospraying technique as well as the physical understanding of the droplet evolution in the electrospraying process [21-24]. The trajectories of droplets are affected by many controllable parameters, i.e. syringe feed rate, concentration of solution, applied voltage and needle gauge. The effects of those controlled parameters on size and morphology of droplets were previously studied [5,6,24-26]. The size of the electrosprayed droplets was able to be tuned by adjusting controlled parameters [5,6]

However, the effects of controllable parameters on deposition area of electrosprayed droplets has not been reported yet. The diameter of the electrosprayed area ranges from several millimeters to tens of centimeters under different controllable parameters [23]. The delicate tuning of the deposition area can greatly contribute to saving materials. When the size of the deposition area fits well with the targeted substrate, the electrosprayed droplets can be efficiently utilized. Vice versa, when the size of deposition is smaller or larger than the area of the targeted substrate, a full utilization of substrate or the electrosprayed droplets cannot be achieved. In this work, the electrosprayed area of ethanol was studied. The ethanol has been reported as a commonly-used solvent for the preparation of the electrosprayed precursor [2,3,9,11-13]. A simplified two-dimensional model has been developed to study the trajectories of droplets and their deposition area during the electrospraying process. The modeling results are validated by the experimental data.

## 2. Experimental approach

The electrospraying process is illustrated in Fig. 1. As a high voltage is applied to the needle nozzle of a syringe, the liquid surface at the tip of the nozzle quickly forms a pointed cone shape. Because the surface tension pulls the liquid back to the nozzle, and Coulomb repulsive force drives the liquid towards the grounded collector. This cone is called "Taylor cone" [21]. Once the surface tension is overcome by Coulomb repulsive force, the liquid jet is then emitted through the apex of the Taylor cone. Eventually, the highly charged liquid breaks into small droplets.

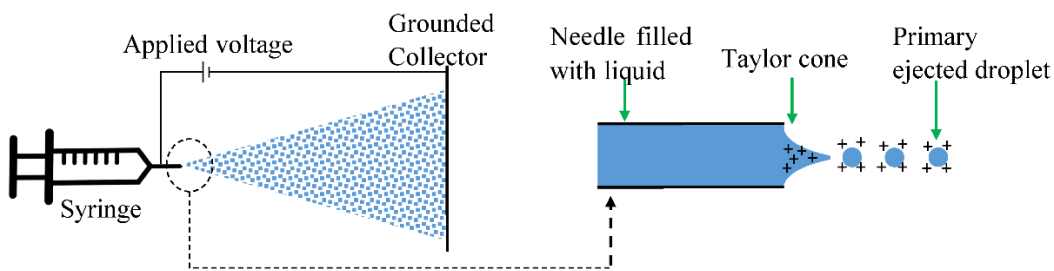

Figure 1: Schematic of electrospraying process.

*2.1 Materials*

Ethanol (200 proof) was purchased from Decon Laboratories Inc., USA. Methylene blue (MB) aqueous solution with a concentration of 1.5 w/v % (1.5 g/ 100 mL) was purchased from Sigma-Aldrich, USA. All the materials were used as received without further purifications. Methylene blue was used to dye the ethanol solution for direct observations of electrosprayed droplets on target. Two drops of methylene blue were added into 25 mL ethanol. Syringe and needle were purchased from PrecisionGlide.

*2.2 Electrospraying process*

The ethanol-MB solution was electrosprayed as shown in Fig. 2. The parameters used were: 5, 7.5, 10, and 12.5 kV for applied voltages; 0.1, 0.5, 1, 2, and 5 mL h$^{-1}$ for feed rates; 1, 3, 5 and 10 cm for the distance between the tip and collector. The inner and outer diameter of the needle is 0.337 mm and 0.641 mm, respectively. The nozzle of the needle was ground with a sand paper and finished with a blunt tip. The collector for electrosprayed droplets was a metal plate covered with a white paper sheet. The metal plate (length: 550 mm, width: 340 mm and thickness: 2 mm) was grounded during electrospraying. The environmental temperature of working condition was 20 °C.

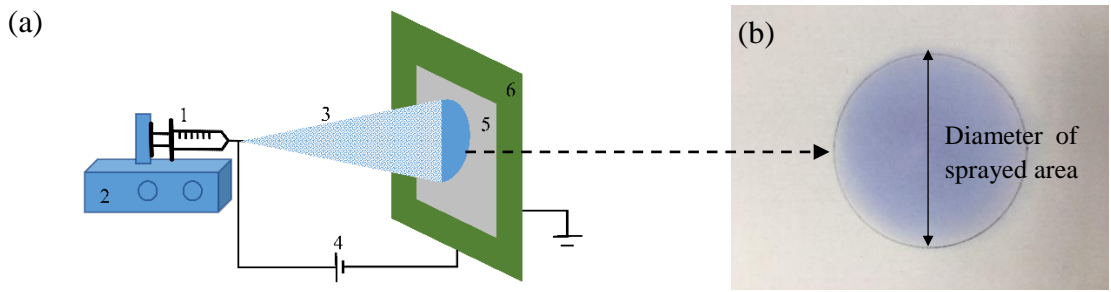

Figure 2: (a) Schematic of electrospraying setup: 1-syringe; 2-feed rate controller; 3-electrosprayed droplets; 4-applied voltage; 5-white paper sheet and 6-grounded metal plate and (b) Typical ethanol deposition area after electrospraying.

## 3. Modeling approach

A simplified two-dimensional model is developed to simulate the trajectories of electrosprayed droplets. It is assumed that the electrosprayed droplets are ejected from the tip of Taylor cone with a velocity in a random direction towards the collector, and all the droplets are spherical and identical in size with predefined parameters during the electrospraying process. The evaporations of droplets are not considered in this model. The computational configuration of electrospraying consists of three domains as shown in Fig. 3: syringe needle (#1 domain), metal collector (#2 domain) and air region (#3 domain). The needle has a length of 20 mm and a width of 0.5 mm. The tip of the needle has a wedge angle of 100°, which simulates Taylor cone in the cone-jet mode of electrospraying process [21]. Electrosprayed liquid is assumed to be fully stored in the syringe needle. A high voltage (5 – 12.5 kV) is applied on the top and bottom side of the needle. The droplets are ejected from the tip of the needle. The metal collector is 340 mm long and 5 mm wide, and four sides of the collector are grounded. The #3 domain is 500 × 500 mm and filled with air.

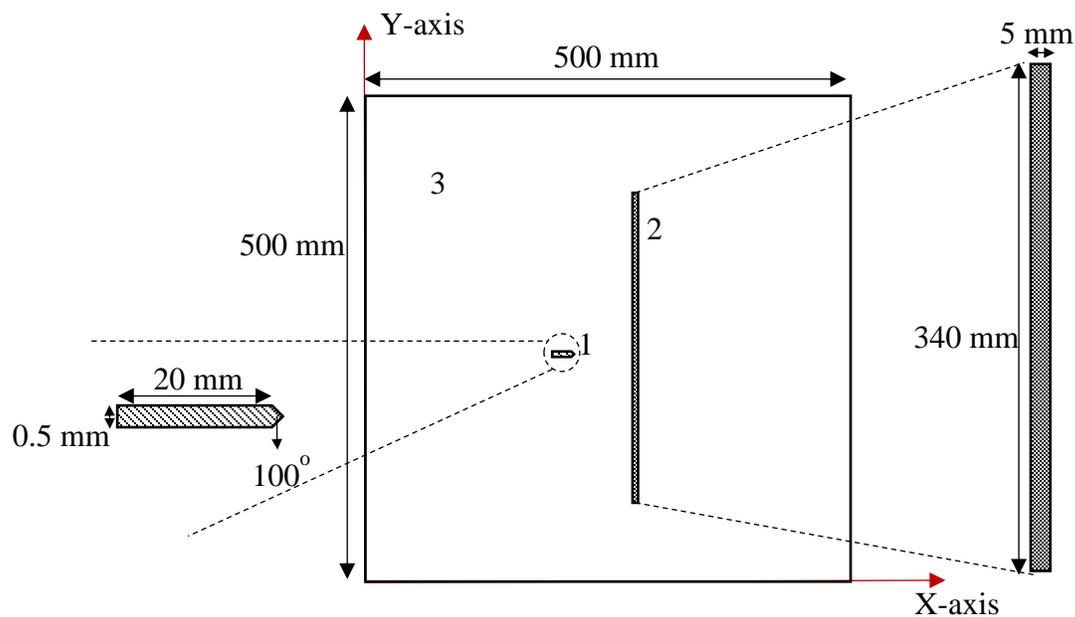

Figure 3: A simplified two-dimensional model with three domains of electrospraying process: 1-syringe needle (#1 domain); 2-metal collector (#2 domain) and 3-air region (#3 domain).

## 3.1 Force scaling analysis

In the movement from the needle tip to the collector, the electrosprayed droplets in the air are subjected to five possible forces: electric force ($F_E$), drag force ($F_D$), Coulomb repulsive force ($F_C$), gravitational force ($F_G$) and buoyance force ($F_B$) described as shown in Fig. 4. Force scaling analysis can capture dominant forces involved in droplet dynamics through the whole process of electrospraying and simplified the model.

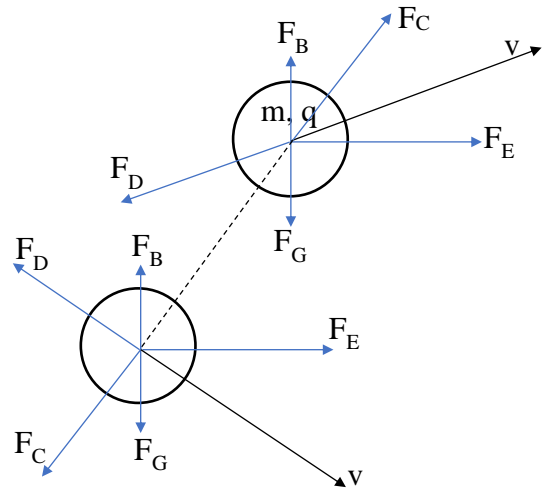

Figure 4: A sketch of two droplets subjected to five possible types of forces.

The formulas for the five forces are given:

$$F_E = qE \tag{1}$$

$$F_D = 3\pi \mu_{air} v d \tag{2}$$

$$F_C = \frac{q^2}{4\pi\varepsilon_0} \cdot \sum_{j=1}^{N} \frac{r-r_j}{|r-r_j|^3} \tag{3}$$

$$F_G = mg \tag{4}$$

$$F_B = -\rho_{air}Vg \tag{5}$$

Where, $q$ is the charge of the droplet, $E$ is the intensity of electric field, $\mu_{air}$ is the dynamic viscosity of the air, $v$ is the velocity of the droplet, $d$ is the diameter of the droplet, $\varepsilon_0$ is the vacuum permittivity, $r$ is the position of the droplet, $r_j$ is the position of any $j$ droplet, $m$ is the mass, $V$ is the volume of the droplet and $\rho_{air}$ is the density of air. The force scaling analysis was made to estimate the magnitude of each force. The magnitude ranges of five forces are estimated using the equations (1) – (5) and summarized in Table 1.

Table 1: Results of force scaling analysis of five forces during electrospraying: The estimated velocity of droplets ranges from 1 to 40 m s$^{-1}$; the distance between the tip and collector ranges from 1 to 10 cm; the applied voltage ranges from 5 to 12.5 kV and the assumed spacing between two droplets is 50 μm.

| Droplet diameter (μm) <br> Force (N) | 2 | 10 | 20 |
|---|---|---|---|
| $F_E$ (Electric force) | 5.52×10$^{-10}$ <br> – 1.38×10$^{-8}$ | 6.17×10$^{-9}$ <br> – 1.54×10$^{-7}$ | 1.76×10$^{-8}$ <br> – 4.39×10$^{-7}$ |
| $F_D$ (Drag force) | 3.40×10$^{-10}$ <br> – 1.36×10$^{-8}$ | 1.71×10$^{-9}$ <br> – 6.82×10$^{-8}$ | 3.41×10$^{-9}$ <br> – 1.36×10$^{-7}$ |
| $F_C$ (Coulomb repulsive force) | 4.36×10$^{-10}$ | 5.48×10$^{-8}$ | 4.40×10$^{-7}$ |
| $F_G$ (Gravitational force) | 3.30×10$^{-14}$ | 4.13×10$^{-12}$ | 3.30×10$^{-11}$ |

| $F_B$ (Buoyance force) | 5.12×10⁻¹⁷ | 6.40×10⁻¹⁵ | 5.12×10⁻¹⁴ |

It can be seen that the electric force, drag force and Coulomb repulsive force are the three main forces that affect the trajectories of droplets' motion through scaling analysis. Gravitational force and buoyance force are much smaller compared with the electric force and drag force, and it is about five to eight orders of magnitude smaller. Therefore, gravitational force and buoyance force contribute much smaller to the droplets' trajectories and will be ignored. Only electric force, drag force and Coulomb repulsive force are considered in this model. During the electrospraying, electric force accelerates droplets towards the collector, while drag force retards the movement of droplets in the opposite direction. Since the droplets are small in size and mass, electric force and drag force quickly reach a balance. Coulomb repulsive force is strongly affected by the distance between two droplets. The ejected droplets near the tip will strongly drive each other far away.

*3.2 Boundary and initial conditions*

The trajectories of ejected droplets subjected to three main forces: electric force, drag force and Coulomb repulsive force from needle tip to collector are simulated using COMSOL Multiphysics software package, which has been demonstrated as an effective tool to study various transport phenomena by others' work [27-30]. This model couples laminar flow module, electrostatics module and particle tracing module. The relative tolerance used is $5\times10^{-5}$. The boundary conditions defined are summarized in Table 2. The mesh independent analysis is also studied. The mesh number of 5,946, 9,130, 15,080 and 33,494 yield the same results. Thus, 5,946 elements are used for the simulation results discussed below.

Table 2: Boundary conditions.

| Boundaries | Conditions | Notation |
|---|---|---|
| #1, #2 | Applied voltage and no slip wall | 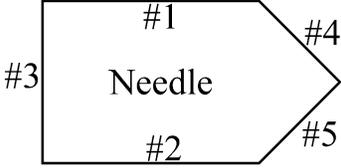 |
| #3, #4, #5 | No slip wall | |
| #6, #7, #8, #9 | Grounded and no slip wall | 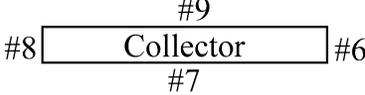 |
| #10, #11, #12, #13 | Zero charge and no slip wall | 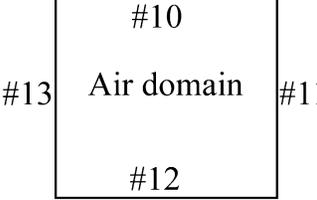 |

The diameter of the primary ejected droplets was estimated by the scaling law [26]. For a given liquid, the diameter of the ejected droplets depends on the feed rate of the liquid. For liquids with very low conductivity and viscosity ($\delta = \left(\frac{\epsilon_0^2 \gamma^3}{K^2 \mu_{liq}^3 Q}\right)^{\frac{1}{3}} \gg 1$) [26], the diameter of the droplets can be estimated as

$$\frac{d}{d_0} = 1.2 \left(\frac{Q}{Q_0}\right)^{\frac{1}{2}} - 0.3 \tag{6}$$

$$d_0 = \left(\frac{\gamma \epsilon_0^2}{\rho K^2}\right)^{\frac{1}{3}} \tag{7}$$

$$Q_0 = \frac{\gamma \epsilon_0}{\rho K} \tag{8}$$

Where, $d$ is the estimated diameter of primary ejected droplet, $Q$ is the feed rate, $\gamma$ is the surface tension of the liquid, $\varepsilon_0$ is vacuum permittivity, $\rho$ is the density of the liquid, $K$ is the electrical conductivity of the liquid and $\mu_{liq}$ is the dynamic viscosity of liquid. For liquids with high enough conductivities and viscosities ($\delta \ll 1$) [26], the diameter of the droplets can be estimated as

$$\frac{d}{(\beta-1)^{1/3} d_0} = 1.6 \left[\frac{Q}{(\beta-1)^{1/2} Q_0}\right]^{\frac{1}{3}} - 1.0 \tag{9}$$

Where, $\beta$ is the relative permittivity of the liquid. In addition, in order to calculate the electric force as describe in Eq. (1) and Coulomb repulsive force as described in Eq. (3), the electric charge that a droplet holds can be determined by the Rayleigh limit [21]

$$q = \pi \sqrt{8 \varepsilon_0 \gamma d^3} \tag{10}$$

The initial velocity of the primary droplet is determined with an assumption that the feed rate is conserved in the liquid flow [22,23]

$$v_{initial} = \frac{Q}{\frac{1}{4}\pi \left(\frac{d}{1.89}\right)^2} \tag{11}$$

The droplets are released one after another in a sequence, and the release rate ($R$) of a droplet from the needle tip is based on the ratio of feed rate and droplet size and yields

$$R = \frac{6Q}{\pi d^3} \tag{12}$$

## 3.3 Parameters of droplets' properties

The parameters used in the electrospraying simulation are summarized in Table 3. The droplets' properties with different electrospraying parameters are summarized in Table 4.

Table 3: Parameters of ethanol and air used in the electrospraying simulation.

| Parameters | Symbols | Value | Unit | References |
|---|---|---|---|---|
| Surface tension of ethanol and air | $\gamma$ | 22.1 | mN m$^{-1}$ | [31] |
| Conductivity of ethanol | $K$ | 0.4 | µS cm$^{-1}$ | [32] |
| Density of ethanol | $\rho$ | 789 | kg m$^{-3}$ | |
| Viscosity of ethanol | $\mu_{liq}$ | 1.083×10$^{-3}$ | Pa·s | [33] |
| Relative permittivity of ethanol | $\beta$ | 25 | | [34] |
| Vacuum permittivity | $\varepsilon_0$ | 8.85×10$^{-12}$ | F m$^{-1}$ | |
| Viscosity of air | $\mu_{air}$ | 1.78×10$^{-5}$ | Pa·s | [35] |

Table 4: Properties of droplets at different feed rates.

| Feed rate (mL h$^{-1}$) | Diameter of primary droplet (µm) | Mass of droplet (kg) | Number of elementary charge carried by a droplet | Initial velocity (m s$^{-1}$) | Number of droplets released in 1 ms |
|---|---|---|---|---|---|
| 0.1 | 2.49 | 6.37×10$^{-15}$ | 96,468 | 20.39 | 3,439 |
| 0.5 | 5.98 | 8.83×10$^{-14}$ | 358,985 | 17.68 | 1,242 |
| 1 | 8.59 | 2.62×10$^{-13}$ | 618,671 | 17.11 | 836 |
| 2 | 12.29 | 7.67×10$^{-13}$ | 1,058,207 | 16.73 | 572 |
| 5 | 19.63 | 3.12×10$^{-12}$ | 2,135,352 | 16.40 | 351 |

## 4. Results and discussion

### *4.1 Trajectories of electrosprayed droplets*

The trajectories of electrosprayed droplets under a condition of an applied voltage of 10 kV, distance between the tip and collector of 5 cm and feed rate of 1 mL h$^{-1}$ are shown in Fig. 5. A total of 836 droplets are ejected in 1 ms. After nearly 8 ms, all the droplets arrive at the collector. The droplets have the highest velocity immediately after the ejection at the tip of the needle. The velocity decreases as the droplets move towards the collector, because the drag force hinders the movement of droplets. At the first a few millimeters near the tip, the electric field intensity is the highest, which is due to a sharp edge of the needle tip with a high curvature. As a result, droplets are accelerated and the moving direction is mainly controlled by the electric field. After the electric field intensity gradually reduces to a smaller value, and then the drag force begins to play a role. Since the magnitude of the drag force is proportional to droplet's velocity and the force direction is opposite to droplet's moving direction, the velocity of the droplet is quickly reduced. After that, the electric force and drag force maintain a balance all the way until the droplets reach the collector. As the electric field intensity decreases towards the collector, the balanced velocity decreases as well. It is found that the Coulomb repulsive force helps to disperse droplets near the needle tip, because this force is strongly affected by the distance between two droplets. Since droplet cloud is denser near the tip, droplets can extend to a wider range with the existence of the Coulomb repulsive force. Compared with the trajectories of droplets without Coulomb repulsive force as shown in Fig. 6, the diameter of deposition area with the Coulomb repulsive force is 2 cm larger. Besides, it is clearly to conclude from Fig. 5 and Fig. 6 that the charged droplets are distributed separately with the Coulomb repulsive force, and easily form agglomerations without coupling Coulomb repulsive force. This explains one benefit of electrospraying of generating homogeneous and dispersed droplets by electrospraying technique. All of the modeling results below (Fig. 7 to Fig. 12) are simulated with electric force, drag force and Coulomb repulsive force.

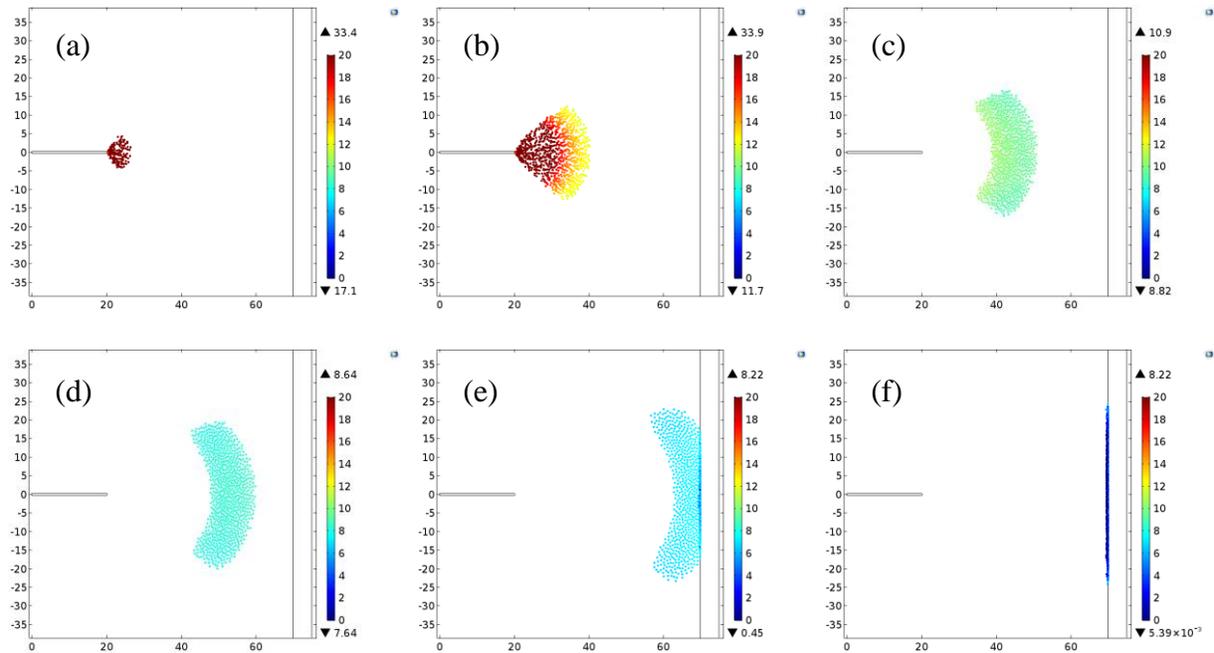

Figure 5: Trajectories of electrosprayed droplets subjected to electric force, drag force and Coulomb repulsive force: (a) t=0.2 ms, (b) t=1 ms, (c) t=2 ms, (d) t=3 ms, (e) t=5 ms and (f) t=7 ms. (Applied voltage: 10 kV, distance between the tip and collector: 5 cm and feed rate: 1 mL h$^{-1}$).

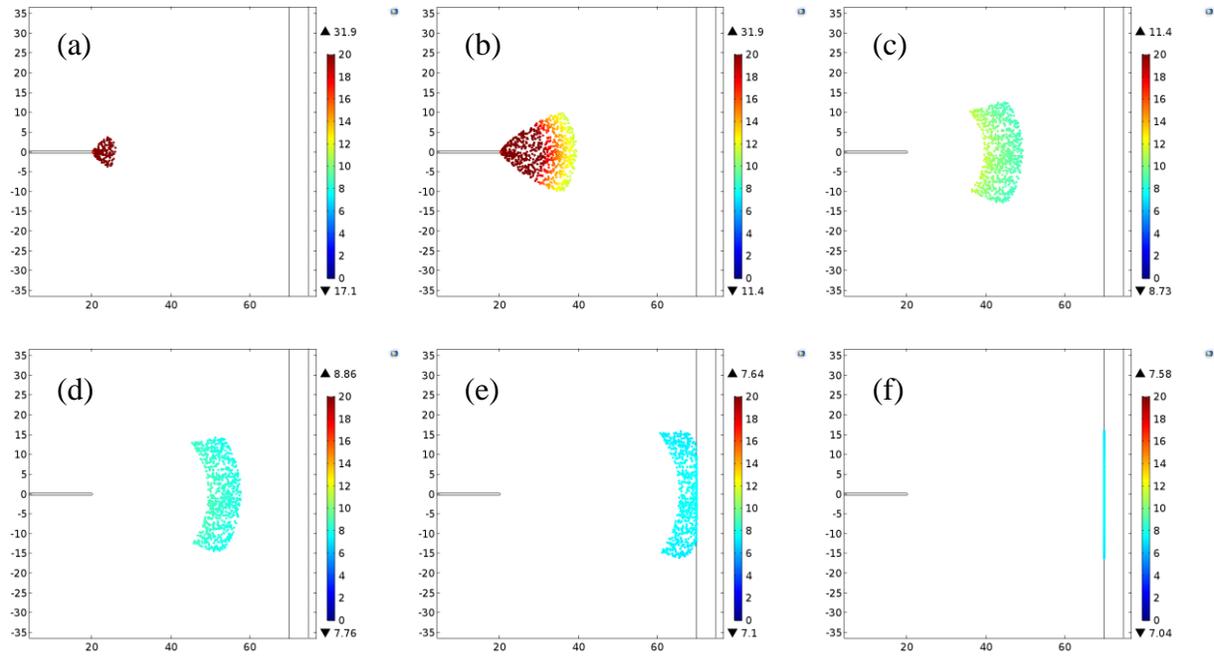

Figure 6: Trajectories of electrosprayed droplets subjected to electric force and drag force: (a) t=0.2 ms, (b) t=1 ms, (c) t=2 ms, (d) t=3 ms, (e) t=5 ms and (f) t=7 ms (Applied voltage: 10 kV, distance between the tip and collector: 5 cm and feed rate: 1 mL h$^{-1}$).

## *4.2 Effect of distance between the tip and collector*

Fig. 7 shows the electric field and deposition area of electrosprayed droplets with a range of distance between the tip and collector from 1 to 10 cm. The controlled parameters used for simulation were 10 kV for applied volatge and 2 mL h$^{-1}$ for feed rate.

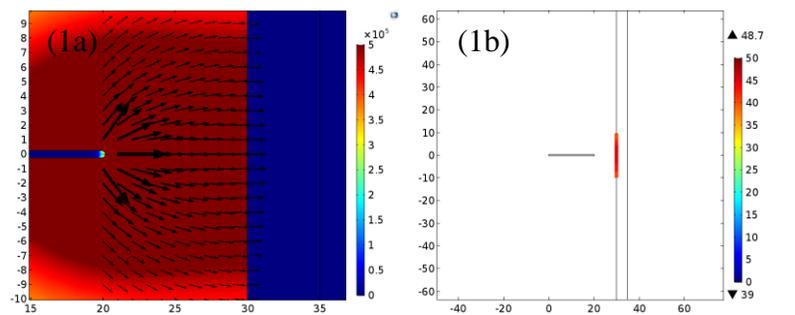

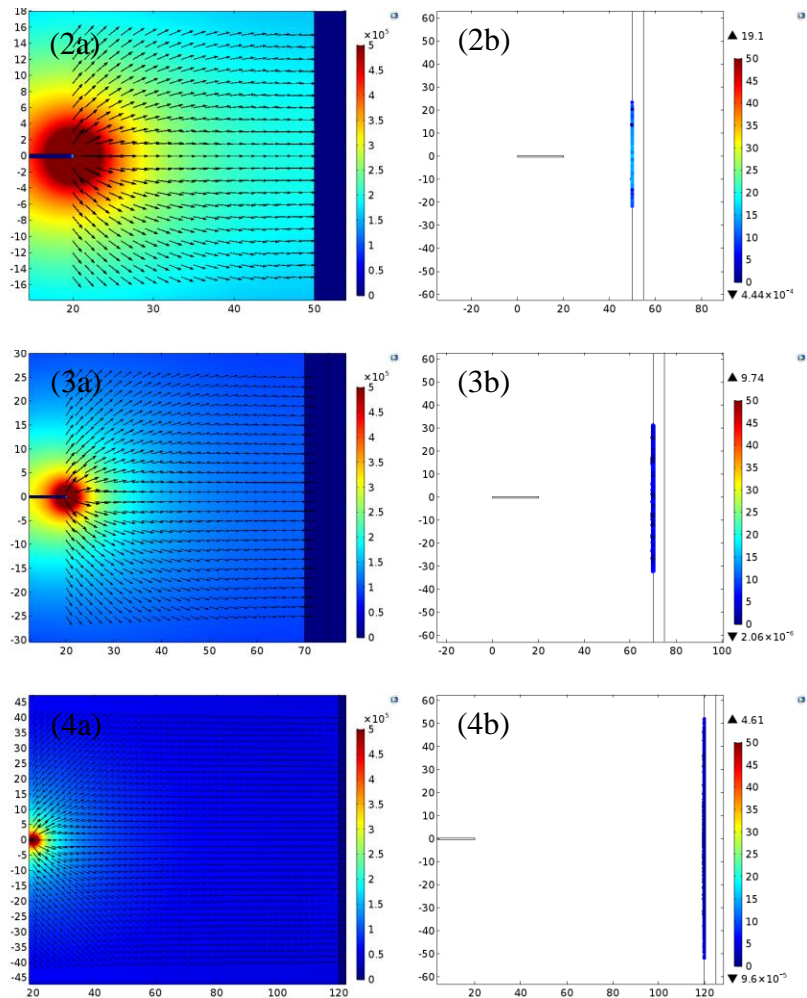

Figure 7: Electric field and deposition area of electrosprayed droplets with different distance between the tip and collector: (1a) and (1b): 1 cm; (2a) and (2b): 3 cm; (3a) and (3b): 5 cm and (4a) and (4b): 10 cm. (1a) – (4a) for the electric field and (1b) – (4b) for the deposition area (Applied voltage: 10 kV and feed rate: 2 mL h$^{-1}$).

The simulated electric field shows that electric field intensity decreases with increased distance between the tip and collector. The average electric field intensity for 1 cm distance is roughly ten times higher than the one for 10 cm distance. The velocity of droplets hence is higher for a short-distance electrospraying. The right column of Fig. 7 explans that the electrosprayed droplets can spread into a larger area with a longer distance between the tip and collecotr. At the initial stage of electrospaying, the droplets are accelarated under the electric force with the *Y*-direction component, and the drag force also increases rapidly with an increased speed. The drag force then reduces the velocity of droplets along the *Y*-direction. At the same time, the electric field line gradully turns from an angle to the vertical direction of the collector, and this means no electric force along the *Y*-direction to balance the drag force. As a result, the velocity of the droplets along the *Y*-direction becomes zero at a later stage. For a short-distance, the electric field line is more denser to the collector, and it results in a smaller deposition area. Fig. 8 illustrates that the diameter of the electrosprayed deposition area always increases with increased distance between the tip and collector with a variety of parameter. Without taking evaporation effect of droplets into account, the deposition area has an approximately linear relationship with distance between the tip and collector.

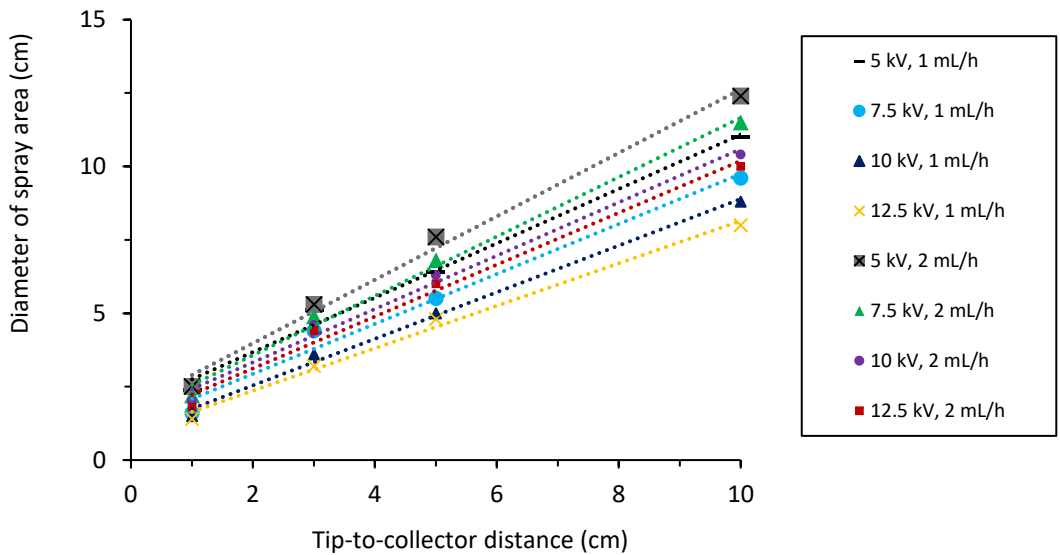

Figure 8: The relationship between the simulated diameter of sprayed area and the distance between the tip and collector.

The simulated diameter of sprayed area and experimental data are compared as shown in Fig. 9. The dashed lines illustrate that the diameter of the electrosprayed deposition area increases with an increased distance between the tip and collector. It can be seen that the simulation results match better with experimental data at a smaller distance between the tip and collector. The discrepancy increases at a larger distance. This may be attributed to the evaporation of liquid droplets. The evaporation effect is not considered in this model. In the experiments, droplets evaporate during the movement towards collector. At a critical stage when the electrostatic force overcomes the surface tension force, then the droplet splits into two or several smaller droplets. At this moment, the smaller droplets have velocities along the *Y*-direction and the trajectories of droplets are further expanded. During the whole process, the droplets can even split for several times in the experiments.

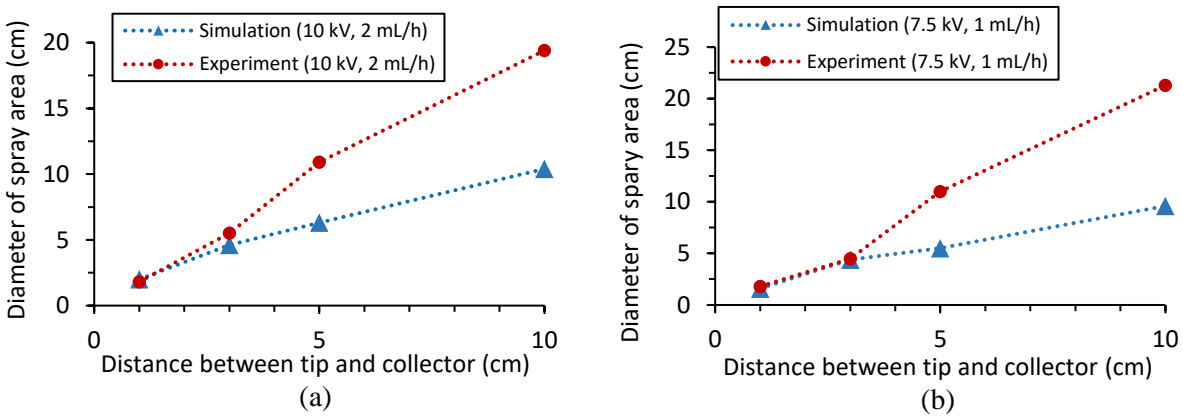

Figure 9: Comparisons between simulated diameter of sprayed area and experimental data with different distance between the tip and collector: (a) Applied voltage: 10 kV and feed rate: 2 mL h$^{-1}$ and (b) applied voltage: 7.5 kV and feed rate: 1 mL h$^{-1}$.

## *4.3 Effect of feed rate*

Fig. 10 and Fig. 11 show the effects of feed rate on diameter of electrosprayed area. It can be seen that the diameter of the deposition area increases with the increased feed rate. It was also reported from previous experiments that a higher feed rate can lead to a larger first-ejected droplet

[6,25,26]. Larger droplets are capable to carry larger electric charges, which induce larger electric force and consequently larger deposition area.

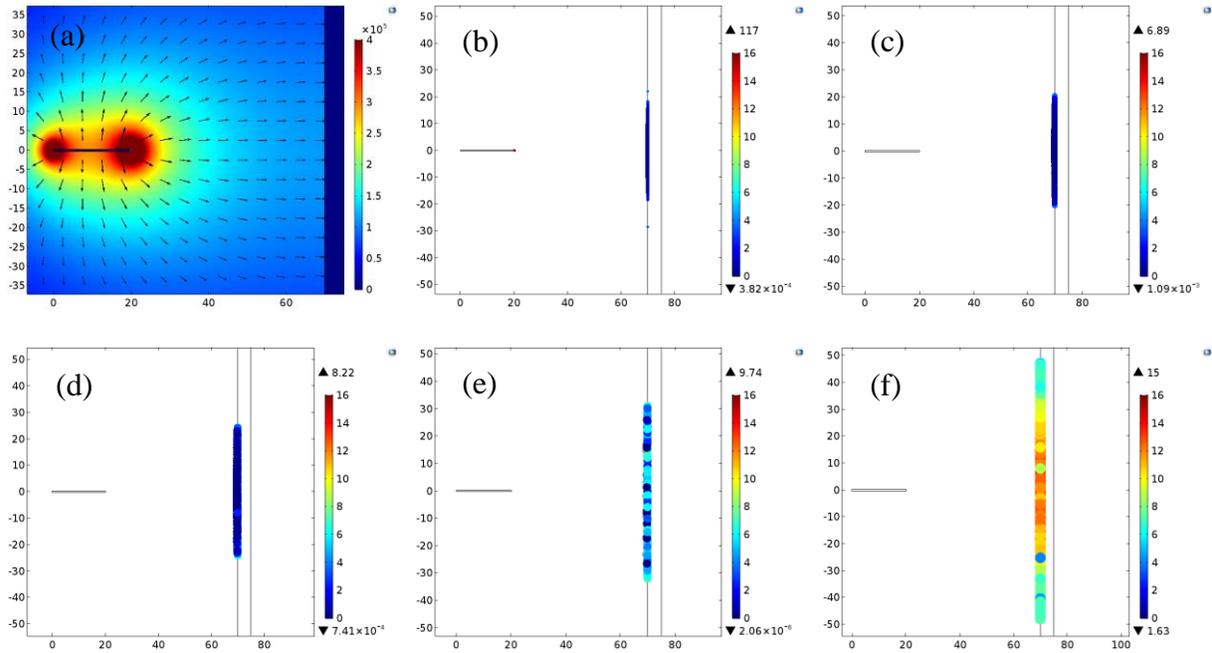

Figure 10: Diameter of sprayed area with different feed rates: (a) electric field and (b) – (f) deposition area: (b) 0.1 mL h$^{-1}$, (c) 0.5 mL h$^{-1}$, (d) 1 mL h$^{-1}$, (e) 2 mL h$^{-1}$ and (f) 5 mL h$^{-1}$ (Applied voltage: 10 kV and distance between the tip and collector: 5 cm).

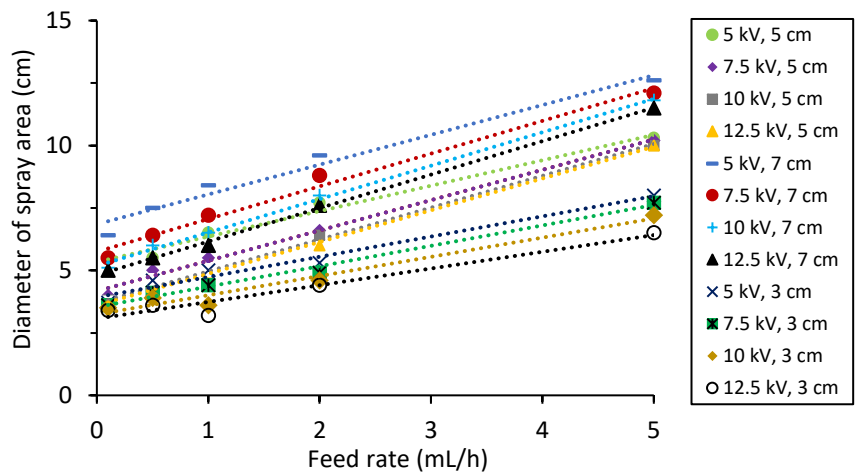

Figure 11: The relationship between the diameter of sprayed area and feed rate under various applied voltages and distances between the tip and collector.

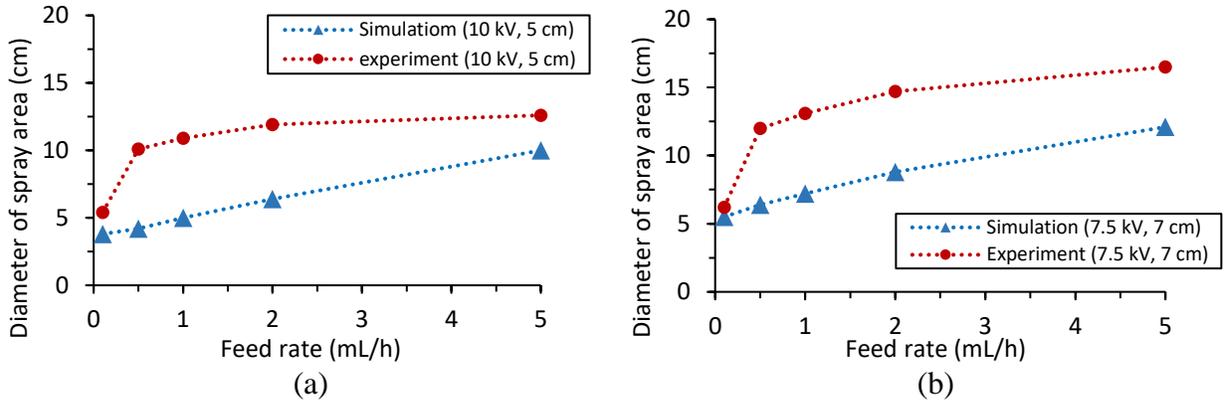

Figure 12: Comparisons of simulation results and experimental data with different feed rates: (a) Applied voltage: 10 kV, distance between the tip and collector: 5 cm and (b) applied voltage: 7.5 kV, distance between the tip and collector: 7 cm.

Fig. 12 compares the experimental and simulated results under two different electrospraying conditions. Both modeling results and experimental data show the same trend that the diameter of deposition area increases with increased feed rate. However, the increasing behavior is slightly different between the modeling and experimental results. The experimental data shows that the diameter of sprayed area increases rapidly with increased feed rate at a lower feed rate range, and increases much slower at a higher feed rate range. This increased characteristic is not demonstrated in the simulations. The discrepancy between the modeling results and experimental data seems to be smaller under a smaller or a higher feed rate. A possible explanation for the discrepancy under a smaller feed rate is concluded. A slower feed rate generates smaller droplets, which have lower evaporation rates. With less evaporations, the simulated results may match better with experimental data at a lower feed rate. Also, there is a possible reason for the decreasing discrepancy under a higher feed rate. A higher feed rate generates larger droplets and the splitting velocity for a larger droplet at the moment when a droplet splits is smaller than the one of a smaller droplet. A smaller splitting velocity will results in a much smaller deposition area. Fig. 13 shows an example of splitting process. Assume two spherical stagnant droplets with a radius of 1 μm and 10 μm, respectively. At the critical moment, they split into two identical spheres. Before splitting, the droplets have electrostatic and surface energies. After splitting, some electrostatic and surface energies are converted into kinetic energy. The total energy is assumed to be constant during the splitting process, and the splitting velocity is given [22]

$$\frac{1}{2} \cdot \frac{q^2}{4\pi\varepsilon_0 r_1} + 4\pi r_1^2 \gamma = 2 \times \left( \frac{1}{2} \cdot \frac{(\frac{q}{2})^2}{4\pi\varepsilon_0 r_2} + 4\pi r_2^2 \gamma + \frac{1}{4} m v_{splitting}^2 \right) \qquad (13)$$

Where, $r_1$ is the radius of a droplet before splitting and $r_2$ is the radius of a droplet after splitting. For a much smaller stagnant ethanol sphere ($r_1$ = 1 μm), the splitting velocity of the droplet is estimated to be 8.95 m s$^{-1}$ and the splitting velocity is estimated to be 2.83 m s$^{-1}$ for the larger stagnant ethanol sphere ($r_1$ = 10 μm). Thus, for a higher feed rate, the ability of increasing the deposition area may dramatically decrease due to a smaller splitting velocity and this may explain a smaller discrepancy between the modeling results, which do not consider the splitting process, and experimental data for a higher feed rate.

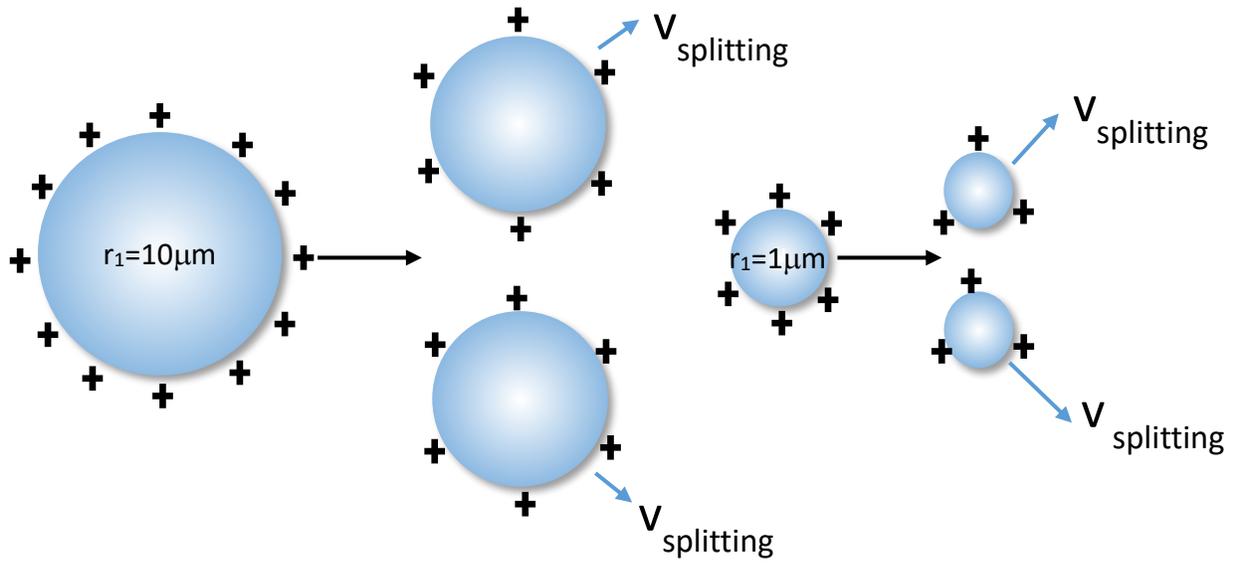

Figure 13: A proposed schematic of droplet splitting: large and small droplets.

## Conclusions

A simplified two-dimensional model was developed to study droplet dynamics and deposition area during the electrospraying process. The deposition area of electrosprayed droplets is tunable with various controllable parameters. It was found that the diameter of electrosprayed area increases with an increased feed rate, and an increased distance between the needle tip and collector for both modeling and experimental results. The discrepancy between modeling and experimental data may be due to the evaporation and splitting, which are not considered in the model. This fundamental understanding should contribute to the optimized choice of controllable parameters for the efficient utilization of electrosparyed droplets or substrate in reality.

## Acknowledgements

This work is supported by US National Science Foundation with an award number: 1563238.